\documentclass[12pt,osajnl2,showpacs,superscriptaddress]{revtex4}  
\usepackage{hyperref} 
\begin{document}

\title{Measurement of Thermo-Elastic Deformation of an Optic using a Polarization Based Shearing Interferometer}

\author{Peter Beyersdorf}
\author{Mark Cordier}
\affiliation{San Jose State University, One Washington Square, San Jose, CA 95192-0106 USA}
\email{peter.beyersdorf@sjsu.edu}

\begin{abstract}A shearing interferometer is presented which uses polarization control to shear the wavefront and to modulate the interference pattern.  The shear is generated by spatial walk-off in a birefringent crystal.  By adjusting the orientation of the birefringent crystal, the components of the wavefront gradient can be independently measured to allow determination of the full wavefront vector gradient as well as reconstruction of the wavefront.  Further, the monolithic nature of the crystal used for shearing allows the interferometer to be setup without  need for precise alignment of any components.  An algorithm incorporating homodyne detection is presented which analyzes the modulated interferograms to determine the components of the wavefront gradient, from which the wavefront is reconstructed.  The thermal deformation of a mirror subject to heating from absorption of a Gaussian pump beam was accurately observed with a sensitivity better than $\lambda/160$.  We show that this sensitivity is scale invariant, and present a method to account for the non-uniform spatial frequency response of the interferometer.
\end{abstract}

\maketitle

\section{Introduction}
Shearing interferometry is a method of common path interferometry whereby a wavefront is interfered with an image of itself that is laterally displaced by the amount of shear, $s$.  The resulting interference pattern occurs only in the region of overlap and yields information about the component of the wavefront gradient in the direction of the shear.   Shearing interferometers can be categorized into one of two type: separate path and near common path.  In the first type a Michelson or Mach Zehnder interferometer is typically used to generate the shear between the interfering beams, while the near common path types typically use glass wedges, etalons or prisms to generate a shear.  The near common path configurations have the advantage of mechanical stability.  The typical challenge associated with using shearing interferometers is interpreting the interferograms which are a function of the wavefront gradient, not displacement \cite{Kothiyal:85,Tuan:08}. 

The interferometer presented here uses spatial walk-off in an anisotropic crystal to generate the wavefront shear without introducing a free-space path difference, so that the common-mode noise rejection of the shearing interferometer is maintained -- our instrument produced stable interference fringes even while pounding on the optical table with a fist.  Further this method does not require critical alignment of optics making it uncomplicated to set-up.  Modulation of the interference pattern by an electrooptic modulator allows the interference pattern to be differentiated from background light, further improving sensitivity.  This is functionally similar to polarization based interferometers developed for shearography\cite{Murukeshan1998, rosso:2006} but we measure the interference for shear in both x and y allowing the full vector gradient to be measured allowing reconstruction of the wavefront.

Measurement of the absolute wavefront profile is possible, but comparison to a reference yields a differential measurement that allows the contribution to the wavefront distortion from a particular element to be isolated.  Here we present a measurement of the thermal deformation of a mirror partially absorbing a Gaussian pump-beam, by measuring the difference in the wavefront with and without the pump beam illuminating the test optic.

\section{Polarization Based Shearing Interferometery}
The interference condition for shearing interferometry is obtained in the usual way for any two-beam interferometer by determining the irradiance at a point due to the electric field from the interfering beams (in this case the original and laterally displaced wavefronts).  In the paraxial approximation, where we assume nearly flat wavefronts, the detected intensity can be expressed in terms of the gradient of the phase front and some uniform phase difference $\Gamma$ that in our experiment comes from the birefringence of the optics:
\begin{eqnarray}\label{eq:Ip}
 I&=&I_1+I_2 + 2 \sqrt{I_1 I_2} \cos{[(\nabla \phi\cdot \vec{s}) + \Gamma]} \
\end{eqnarray}

\begin{figure}[h]
\center
\includegraphics[width=.55\textwidth]{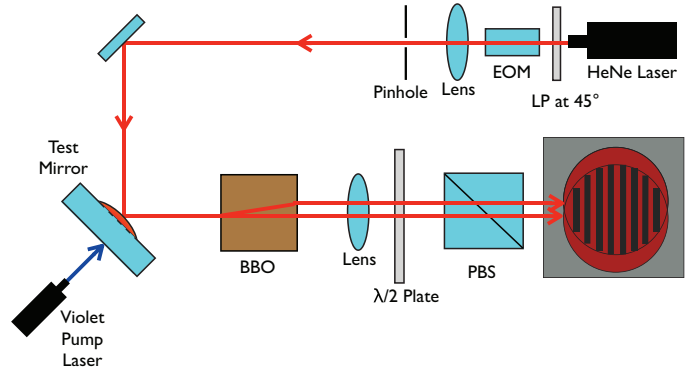}
\caption{\label{fig:experiment}
Schematic diagram of the shearing interferometer.  A Helium Neon laser polarized at $45^{\circ}$ by a linear polarizer (LP) passes through a lens and a $5\;\mathrm{\mu m}$ pinhole to clean the spatial mode of the beam.  The laser then reflects off the test optic.  A BBO crystal shears the beam in x or y depending on its (adjustable) orientation. A half wave plate placed after the BBO crystal rotates the polarization so each component of the sheared wavefront has a polarization component along the transmission axis of the polarizer (PBS).  The interference pattern is recorded by a digital CMOS camera and subsequently analyzed.}
\end{figure}

We use Beta Barium borate (BaB$_2$O$_4$, hereafter `BBO'), a  negative uniaxial birefringent crystal, to laterally shear the two wavefronts in our experimental set-up (shown in Figure~\ref{fig:experiment}).   The shear between the wavefronts is due to the spatial ``walk off'' of the eigenpolarizations of the crystal defined by the angle $\rho$ between the displacement vector $\vec{D}$ and the electric field vector $\vec{E}$ \cite{Fowels} which obey

 \begin{equation}\label{eq:DispVec}
\left[ {\begin{array}{c}
 D_x\\
 D_y\\
 D_z\\
 \end{array} } \right]
 = 
 \epsilon_0 \left[ {\begin{array}{ccc}
 n_o^2 & 0 & 0\\
0 & n_o^2 & 0\\
0 & 0 & n_e^2\\
 \end{array} } \right]
\left[ {\begin{array}{c}
 E_x\\
 E_y\\
 E_z\\
  \end{array} } \right].
 \end{equation} 

The BBO crystal used here is cut with its optical axis at 28$^\circ$ from the surface normal.  The ordinary wave is polarized along the crystal's y-axis, so $D_y=\epsilon_0 n_o^2 E_y$, while the extraordinary wave (in the $x$-$z$ plane) obeys
 \begin{eqnarray}
 \label{eq:DE}
  D\left[ {\begin{array}{c}
 \cos\theta\\
 \sin\theta\\
 \end{array} } \right]
&=&
 \epsilon_0 \left[ {\begin{array}{ccc}
 n_o^2  & 0\\
0 &  n_e^2\\
 \end{array} } \right]
\left[ {\begin{array}{c}
 E_x\\
 E_z\\
  \end{array} } \right].
 \end{eqnarray} 
 The walk off angle is equal to the angle between $\vec{D}$ and $\vec{E}$ for the extraordinary wave and is found by solving 
   \begin{equation}
   \vec{D} \cdot \vec{E} =|D||E| \cos\rho
   \end{equation}
by using equation~\ref{eq:DE}, giving
\begin{eqnarray}
\rho=\cos^{-1}\left( \frac{\left[ \frac{\cos^2\theta}{n_o^2} + \frac{\sin^2\theta}{n_e^2}\right]
}{ \sqrt{\frac{\cos^2\theta}{n_o^4} + \frac{\sin^2\theta}{n_e^4}}}\right).
 \end{eqnarray}
At $\lambda = 632.8$ nm,  the ordinary index of refraction $n_o$  for BBO is $1.6673$ and the extraordinary index of refraction $n_e$ is $1.5500$ \cite{DNN}.  The  displacement between the two beams after traveling through our crystal whose thickness is $L= $13 mm, results in a calculated shear $s=L \tan(\rho)= $ 820 $\mu$m.  The measured value of 860 $\mu$m obtained by measuring the distance between the two shadows (one for each polarization) cast by a thin wire placed just before the BBO crystal, agrees to within 5\% of the calculated result.
\begin{figure}[h] `
\center
\includegraphics[width=.7\textwidth]{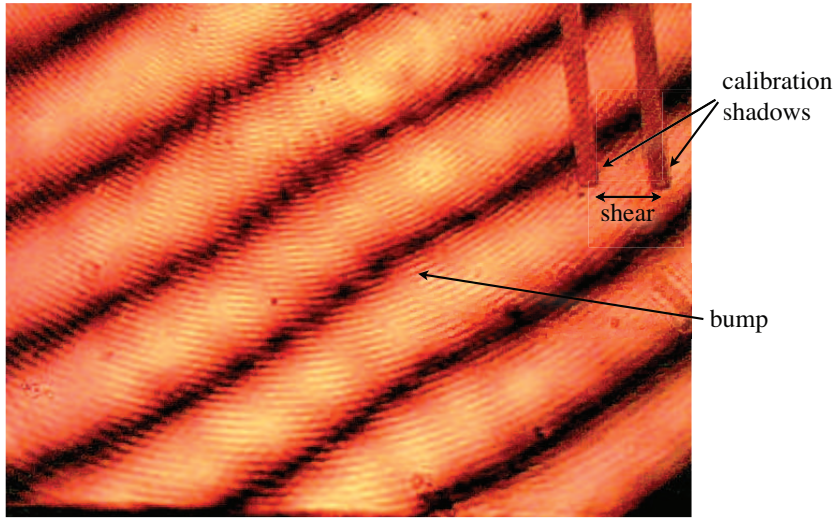}
\caption{\label{fig:interference} An interferogram from the shearing interferometer showing a horizontal shear, as noted by the displacement of one of the two shadows of a pin.  The wavefront being measured has a roughly uniform curvature (responsible for the parallel fringes) and an additional small bump from a thermally deformed optic.  The effect of the bump is visually evident as a deviation in the straightness of the fringes.}
\end{figure}

The interference pattern that results from the shearing interferometer is superimposed on any background light that illuminates the imaging sensor.  To separate the the interference pattern from the background light in the detected image, the relative phase between the interfering beams $\Gamma$ is sinusoidally modulated.  This is accomplished by using a z-cut Lithum Niobate electrooptic modulator placed at the output of the laser, oriented such that the ordinary and extraordinary waves have equal amplitude.   The birefringence of the Lithium Niobate crystal depends on the voltage, $V$, applied to the electrodes on the z-faces and is
\begin{equation}\Gamma=\frac{2 \pi}{\lambda} (n_e-n_o) L-\frac{\pi}{\lambda} (n_e^3 r_{33} - n_o^3 r_{13}) \frac{L}{d} V\end{equation} 
where the crystal length is $L$, the electrode separation is $d$ and $n_e$ and $n_o$ are the extraordinary and ordinary indices of refraction respectively and $r_{13}$ and $r_{33}$ are elements of the electrooptic tensor \cite{Yariv&Yeh}.  For our geometry $d=4\;\mathrm{mm}$, $L=25\;\mathrm{mm}$ which gives a half-wave voltage of about 450 V.  We drive our modulator with a 150 V sinusoidal signal at about 0.3 Hz well below the Nyquist frequency of 7.5 Hz for our measurement.  The modulated interferogram is demodulated in software by our interferogram processing algorithm.

In order to reconstruct the curvature of the wavefront in two dimensions the interferogram is recorded with one orientation of the BBO crystal, the crystal orientation is then rotated by 90 degrees around the propagation axis of the beam to record the interferogram for the other component of the wavefront gradient.   In principle, a beam splitter may be used to illuminate a pair of orthogonally oriented crystals, allowing simultaneous measurements of both gradient components.  To ensure that the crystal rotation did not introduce unwanted translation of the beam, the shadow of a pin placed before the crystal in the field of view of the camera is observed to remain fixed (for the ordinary wave), although a small amount of translation of the beam relative to the shear could be tolerated, since the shear determines the resolution of the wavefront reconstruction. The light was polarized at $45^{\circ}$ in order to have equal components along the principle axes of the crystal.  

The optical path length from the pinhole to the test mirror is 168 cm and from the test mirror to the BBO is an additional 125 cm resulting in an image magnification from the test mirror to the BBO of 1.74x.   We place a thin (250 $\mu m$) wire into the path of the beam just before the BBO and measure the width of the shadow on the image to calibrate the spatial scale of the interferogram, and measure the separation of the (orthogonally polarized) shadows of the pin to determine the shear.   A half wave plate placed after the crystal rotates the polarization once again so that each principle polarization state of the crystal has a component along the transmission axis of a fixed orientation polarizer placed in front of the imaging sensor.  A lens images the front of the BBO crystal onto the imaging sensor with a magnification of -0.45x for a net magnification from the test mirror onto the imaging sensor of -0.79x.

\section{Interferogram Processing and Wavefront Reconstruction}
The distortion of the wavefront is determined by measuring each component (x and y) of the distorted wavefront and comparing each to a reference measurement. 
Each measurement consists of an analysis of the modulated interferogram over 30 seconds of video collected by a consumer level 8-bit color webcam (Agama-V) with a frame-rate of 15 frames per second.  The time series for each  pixel is independently processed. 
\begin{figure}[b]
\center
\includegraphics[width=.75\textwidth]{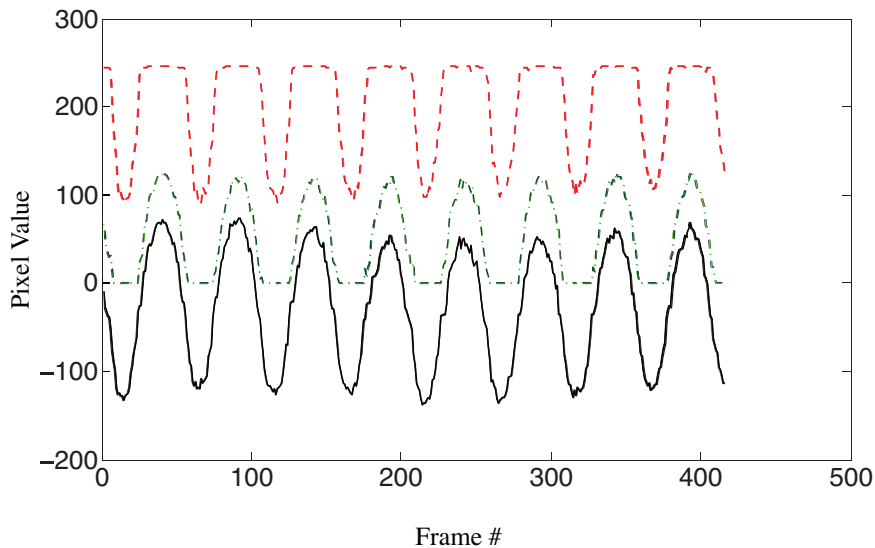}\label{fig:stitch}
\caption{Red (upper dashed curve) and green (lower dashed curve) values for a representative pixel as a function of time.  The composite value (solid line) is formed by stitching their derivatives together after appropriate scaling, then integrating. }
\end{figure}
The effective dynamic range of the webcam is increased by stitching together the data from the red and green channels of each pixel.  The red channel is most sensitive to the 632.8 nm light and for our illumination is often saturated at points with constructive interference.  The green channel is less sensitive and does not saturate, but fails to register a non-zero value for many points with destructive interference.   We stitch the red and green channels together to get a composite data channel as follows:  An appropriate threshold for the illumination level is chosen such that if the green pixel value exceeds this threshold the green data is considered more reliable (since the red value would be at or near saturation), and below this threshold the red data is considered more reliable.  For our illumination, a threshold value of 32 (out of 255 for our 8-bit sensor) was used.  We then take the frame-to-frame difference of each data channel, so it is the derivative of the data being stitched together, avoiding discontinuities in the composite data.  We  empirically determine the appropriate scale factor for the derivative of the green data to compensate for the amplitude difference between the red and green values.  The scale factor is determined by comparing the frame-to-frame difference of the data in the red channel to the frame-to-frame difference of the data in the green channel for frames where the data in the green channel crosses our chosen threshold.  The green derivative data channel is then scaled by this factor and inserted in place of the red derivative data in regions where the green values exceed our chosen threshold.  The cumulative sum of this composite derivative data channel is then computed to give a composite data channel free of discontinuities and with minimal kinks at the stitching points.  An example of the time series of data for a single pixel is shown in Figure~\ref{fig:stitch}.

With the birefringence $\Gamma$ modulated cosinusoidally so that $\Gamma=\Gamma_0 + m \cos( \omega t)$  the interference term in equation \ref{eq:Ip} is 
\begin{equation} 
2 \sqrt{I_1 I_2}\cos\left( \nabla \phi \cdot \vec{s} + \Gamma_0 + m \cos( \omega t) \right).
\end{equation}   Using the sum angle formula for cosines
 \begin{equation} 
\cos\left( \nabla \phi \cdot \vec{s} + \Gamma_0 + m \cos( \omega t) \right)=\cos (\nabla \phi \cdot \vec{s} + \Gamma_0) \cos(m \cos( \omega t) ) - \sin (\nabla \phi \cdot \vec{s} + \Gamma_0) \sin(m \cos( \omega t) )  \end{equation}
along with the Jacobi-Anger expansion allows this to  be expanded as  a series of sinusoidal components with Bessel function amplitudes  \cite{Abramowitz1970}:
\begin{eqnarray}\label{eq:jacobianger}
\cos\left( \nabla \phi \cdot \vec{s} + \Gamma_0 + m \cos( \omega t) \right)&=&\cos (\nabla \phi \cdot \vec{s} + \Gamma_0) \left[ J_0(m) + 2 \sum_{n=1}^\infty (-1)^n J_{2n}(m) \cos(2 n \omega t) \right] \nonumber \\ &-&
\sin (\nabla \phi \cdot \vec{s} + \Gamma_0)\left[ - 2 \sum_{n=1}^\infty (-1)^n  J_{2 n-1}(m) \cos((2 n-1) \omega t) \right].
\end{eqnarray}
The phase shift which depends on the gradient, $\nabla \phi \cdot \vec{s}+\Gamma_0$, can be determined if  the modulation depth $m$ and amplitude $A$ of an even and an odd order frequency harmonic of the modulation frequency $\omega$ are known. Writing the amplitude of the first three harmonic frequency components as
 \begin{eqnarray}\label{eq:jacobiangerN1}
A_1&=& -2 \sin (\nabla \phi \cdot \vec{s} + \Gamma_0) J_{1}(m) \\
 A_2&=&-2 \cos (\nabla \phi \cdot \vec{s} + \Gamma_0) J_{2}(m)\\
 A_3&=&2 \sin (\nabla \phi \cdot \vec{s} + \Gamma_0) J_{3}(m)
\end{eqnarray}
and taking the ratio of the measured harmonic amplitudes $A_1/A_2$ allows us to solve for $\nabla \phi \cdot \vec{s}$, 
\begin{equation}
\label{eq:phitowaverfont}
\nabla \phi \cdot \vec{s} =\tan^{-1} \left( \frac{J_{2}(m) A_{1}}{J_{1}(m) A_{2}}\right)- \Gamma_0.
\end{equation} 

Here $\Gamma_0$ is a constant term that represents any static phase shift due to birefringence of the optics and is irrelevant for our analysis since it cancels out when comparing the measurement of a distorted wavefront to that of a reference.  
In order to shift the time origin so that the modulation waveform is purely cosinusoidal in the form of equation \ref{eq:jacobiangerN1}, we record the complex Fourier transform amplitude and phase of the fundamental, second and third harmonic frequency components for each pixel, and subtract one, two and three times respectively the measured phase angle of the fundamental resulting in a purely real fundamental component - equivalent to time shifting the data so that the modulation is cosinusoidal.
 
The modulation depth and frequency were determined implicitly from the video data.  Ideally the modulation frequency would be synchronized to the frame rate of the camera such that each cycle of modulation corresponds to exactly 6 frames of data.  This would allow an integer number of cycles of the fundamental, second harmonic and third harmonic modulation frequency (those used in the analysis) to be measured without spectral leakage in the discrete Fourier transform, while minimizing the acquisition time.  Because we had no convenient way to synchronize the modulation and acquisition rate we instead used a long integration time to measure many cycles of the modulation and process a subset of this data that has a near integer number of modulation cycles, thus minimizing spectral leakage.   The modulation frequency of 0.30 Hz was found by computing the temporal fast Fourier transform on each pixel and determining the lowest non-zero frequency peak for the mean amplitude of the data for all pixels.  To reduce quantization error and to allow for fast processing of the Fourier transforms the waveforms were first padded with zeros to a length of 1024 points (the recorded data series length was 450 points, truncated a bit below this to allow a near integer number of cycles to be contained in the data).   

The modulation depth is found by analyzing the relative amplitude of the fundamental and third harmonics of the modulation.  The ratio of the amplitude of the fundamental $A_1$ and third harmonic $A_3$ given in equation~\ref{eq:jacobiangerN1} satisfy
\begin{equation}\frac{A_1}{A_3}=-\frac{J_1(m)}{J_3(m)}.\end{equation}
We find the average value of $\frac{A_1}{A_3}$ for all pixels and numerically solve for a modulation depth of $m=1.08$ rad by minimizing the root-mean-squared error when compared to this ratio. 

Using the measured modulation depth and period, the value of $\nabla \phi \cdot \vec{s}$ is calculated for each pixel in the frame.  Because the value is cyclic in $2 \pi$ it must be ``unwrapped'' to give the physical phase front.  This is done using a quality guided 2D unwrapping algorithm described in \cite{Ghiglia}.   

Since the unwrapping process can introduce errors, we first take the difference between the distorted and reference wavefronts and unwrap only the difference, making use of equation~\ref{eq:phitowaverfont} and  the difference formula for arctangents \cite{Abramowitz1970}
\begin{equation}
\tan^{-1} (z_d) -\tan^{-1} (z_r) 
= \tan^{-1}\left( \frac{z_d-z_r}{1+z_d z_r}\right). 
\end{equation}
 
If the wavefront distortion is attributable to the deformation of a mirror and a reference wavefront $\phi_{r}$ free of the effects of the mirror deformation is known,
the components of the surface deformation gradient are found from 
\begin{equation}2 \nabla h= \nabla \phi_d- \nabla \phi_{r}\end{equation}
 and equation  \ref{eq:phitowaverfont} and have their mean value subtracted (equivalent to removing any tilt of the wavefront).  The arrays containing $\partial h(\vec{r})/\partial x$ and $\partial h(\vec{r})/\partial y$  are then shifted by an amount corresponding to the average lateral displacement of the interfering beams,  $-\vec{s}/2$.  Finally the surface profile, $h(\vec{r})$, corresponding to the measured gradient components is found using a Poisson reconstruction with Neumann boundary conditions \cite{Agrawal2005}.
 
\section{Sensitivity and Resolution}
\begin{figure}[h]
\center
\includegraphics[width=.55\textwidth]{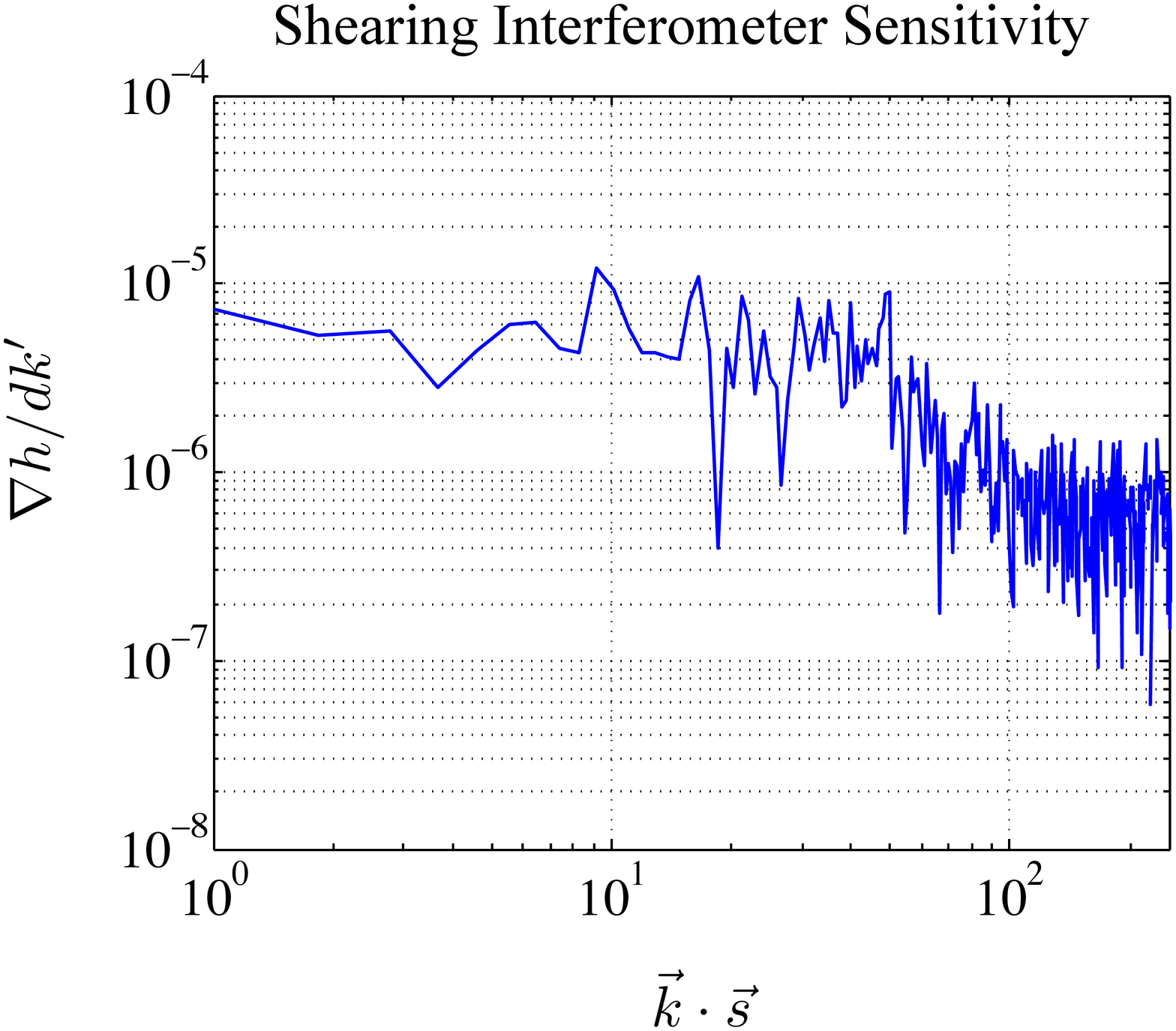}
\caption{
\label{fig:sensitivity}
The measured spectral noise floor of our shearing interferometer.  The spatial frequency units are normalized to the amount of shear so that one spatial frequency unit is $dk'=2 \pi s/ x$ where $s$ is the magnitude of the shear and $x$ is the width of the region of the surface being measured.  For our experiment with $s=860\;\mathrm{\mu m}$ and $x=6.3\;\mathrm{mm}$, $dk'=0.86$ rad.}
\end{figure}

The sensitivity of the shearing interferometer is estimated by the spatial noise spectrum of the processed interferogram shown in  figure \ref{fig:sensitivity}.  The spatial noise spectrum is found from the interferogram recorded for shear in one direction.  The 1D Fourier transform of the central row of pixels is computed and used as a representative noise spectrum.  At the low spatial frequencies associated with a thermal deformation, the average surface gradient noise was $5.8 \times 10^{-6}$ m/m.  We can separate the contribution from the optical magnification and the shear on this noise level by defining $dk'=2 \pi s/ x$ with $s$ the magnitude of the shear and $x$ the width of the region of the surface being measured.  With the spatial scaling effects confined to the term $dk'$ which was 0.86 rad for our experiment, we can express the sensitivity to a  gradient as $5.0 \times 10^{-6}\;\mathrm{(dk')^{{-1}}}$ m/m allowing this result to be scaled to account for the change in sensitivity with different optical magnification and/or shear.  This can be compared to the peak of the gradient calculated from an analytical model.

With the radius and thickness of our mirror being several times greater than the Gaussian radius of the pump beam, the size and shape of the thermo-elastic deformation is well approximated by the analytical solutions for the thermo-elastic deformation in a half-infinte mirror.  For such a geometry, the temperature profile and the resulting magnitude and shape of the thermo-elastic deformation has been worked out\cite{Lu:07} giving a surface displacement of 
\begin{equation}
u_z(r)=\frac{u_c}{8} \left[\mathrm{E}_1 (2 r^2/w^2) + \gamma + \ln(2 r^2/w^2) \right]
\end{equation}
where  $\gamma$ is EulerÕs constant, $\mathrm{E}_1$ is the exponential integral function \cite{Abramowitz1970}, and 
\begin{equation}
u_c=\frac{2 \alpha \epsilon P}{\pi \kappa} (1+\nu)
\label{eq:uc}
\end{equation}
is a characteristic displacement depending only on the illumination power $P$ and material properties, which for our BK7 substrate:  $\kappa=1.114$ W/m K  is the thermal conductivity of the mirror substrate, $\alpha=7.1\times10^{-6}\;\mathrm{K}^{-1}$ is the thermal expansion coefficient, $\epsilon=0.75$ is the fraction of the incident power absorbed, and $\nu=0.206$ is Poisson's ratio.   Defining the difference in surface deformation between a point at the center of the heating beam ($r=0$), and a point at one Gaussian beam radius ($r=w$) as the maximum surface deformation across the beam, 
\begin{equation}
u_{max}\equiv u_{z}(0)-u_{z}(w)
\label{eq:umax}, 
\end{equation}
the maximum gradient from the model of the thermoelastic deformation is $1.44 u_{max}/w$.  With our measured noise level the equivalent displacement sensitivity for thermo-elastic deformations is
\begin{equation}
u_{max}=5.5 \times 10^{-7} \frac{w x}{s}.
\end{equation}
Our measurement uses a 5mm wide CMOS sensor, to observe the test mirror with a magnification of -0.79x, thus the width of the region on the mirror being measured is $x=6.3\;\mathrm{mm}$.  With a value of $s=0.91 w$ we have a displacement sensitivity of 3.8 nm equivalent to better than $\lambda/160$.  
Since the gradient scales inversely with the pump beam radius $w$ while the measured signal scales linearly with the amount of shear $s$, this displacement sensitivity can be achieved for thermal deformations with a larger spatial scale if the size of the shear is also scaled proportionally, limited only by $x$ the field of view of the image.

The theoretical displacement sensitivity limit of this interferometer is limited by the quantization noise of the n-bit imaging sensor.  With the birefringence set to $\Gamma_{0}=\pi/2$, equivalent to biasing the interference to the side of a fringe where the sensitivity is maximized, the minimum detectable value for $A_{2}$ in equation~\ref{eq:phitowaverfont} is $A_{1}/2^{n}$.  The modulation depth produces a value of $\frac{J_{2}(m) }{J_{1}(m)}$ close to unity, so equation~\ref{eq:phitowaverfont} gives
\begin{equation}
\label{eq:phitowaverfontmin}
\nabla \phi \cdot \vec{s} \approx \tan^{-1} \left( 2^{n}\right)- \pi/2 \approx \frac{1}{2^{n}}
\end{equation} 
relating this to the mirror deformation over a distance $s$ using $2 k \Delta h= \nabla \phi \cdot \vec{s}$ gives a displacement sensitivity of 
\begin{equation}
\Delta h_{min}=\frac{\lambda}{4 \pi 2^{n}}
\label{eq:hmin}
\end{equation} 
for our 8-bit camera ($n=8$) the theoretical sensitivity limit evaluates to slightly better than $\Delta h_{min}=\lambda/3000$.

The reconstruction algorithm assumes the gradient of the wavefront can be approximated by the finite difference between points on the wavefront separated by the shear distance $s$,
\begin{equation}\label{eq:finitedifference}\nabla \phi(r) \cdot \vec{s}\approx \phi(\vec{r} + \vec{s}/2)- \phi(\vec{r}- \vec{s}/2).\end{equation}
  This limits the spatial resolution of the instrument to $s$.  If the wavefront $\phi(r)$ can be approximated by a second order Taylor series expansion \begin{equation}\phi(r_0 + s) \approx\phi(r_0) + \phi'(r_0) s + \frac{1}{2} \phi''(r_0) s^2 \end{equation} then the assumptions in expression~\ref{eq:finitedifference} are valid for
\begin{equation}\label{eq:condition}s \ll \frac{2 \phi'}{\phi''}\end{equation}
where $\phi'$ is the gradient of the wavefront in the direction of the shear, and $\phi''/2$ is the curvature in the direction of the shear.  In this regime, the magnitude of the interference term being measured, $\nabla \phi \cdot \vec{s}$, is proportional to $s$, so the sensitivity of the instrument improves linearly with $s$.  Thus if the characteristic length scale, $l_c=\frac{2 \phi'}{\phi''}$, for the deformations being investigated is known, the shear should be approximately equal to $l_c$ to maximize the sensitivity while maintaining sufficient resolution to observe the deformation.  Note that this criteria is less strict that that of expression~\ref{eq:condition}, but with prior knowledge of the expected shape of the wavefront distortion, the tradeoff between sensitivity and resolution can be skewed towards sensitivity with less accuracy measuring the known shape of the distortion due to the filtering of higher spatial frequency components of the distortion.  

Shearing interferometers have a non-uniform spatial frequency response that is well understood \cite{Servin:07}.  The sampling of the phase difference between two points on the wavefront is insensitive to spatial fluctuations with an integer number of cycles within the shear distance.  The data analysis does not attempt to account for this non-uniform spatial frequency response of the shearing interferometer  because of the infinite values in the inverse filter necessary to fully compensate for the frequency response.  Instead we make use of the fact that the surface profile being investigated is  known both through calculation and an independent measurement using a Michelson interferometer, so the effect of the frequency response filter can be calculated and its effect on the measurement of the peak distortion can be used to compensate the measured value of the peak distortion.  

\begin{figure}[h]
\center 
\includegraphics[width=.55\textwidth]{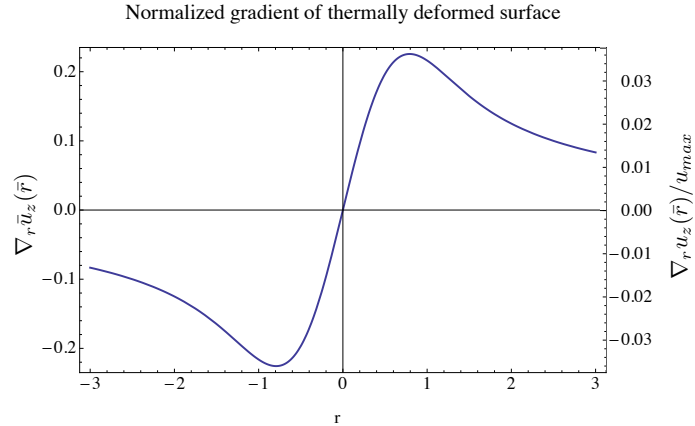}
\caption{\label{fig:normalized_slope}The normalized radial gradient of the longitudinal thermally induced deformation from partial absorption of a Gaussian beam.  The deformation $\bar{u}_{z}=\frac{u_{z}}{u_{c}}$ is normalized to the characteristic thermal deformation $u_{c}$ given in equation~\ref{eq:uc} for the left axis, and normalized to the maximum deformation $u_{max}$ given in equation~\ref{eq:umax} on the right axis.  The transverse coordinate $\bar{r}=\frac{r}{w}$, is normalized to $w$, the Gaussian beam radius of the heating beam}
\end{figure}

\begin{figure}[h]
\center
\includegraphics[width=.55\textwidth]{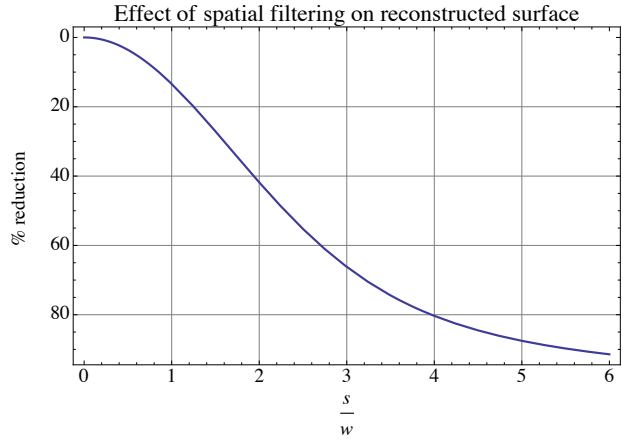}
\caption{ \label{fig:spatial_filtering}The percent reduction in the measured peak surface deformation due to spatial filtering from the shearing interferometer over a spatial length scale $s$.  This is calculated according to expression~\ref{eq:spatialfilteringeffect}  and multiplied by 100\% to express as a percentage.}
\end{figure}

For a wavefront distortion produced by reflection off a thermo-elastically deformed mirror the shape and magnitude of the deformation has been extensively investigated \cite{Hello:90a,Hello:90b,Lu:07,Winkler:91}.  The approximate analytical expression for the thermo-elastic deformation of a mirror heated by absorption of a Gaussian laser beam in \cite{Lu:07} can be differentiated to give an expression for the mirror surface gradient plotted in figure~\ref{fig:normalized_slope}.
From the analytical expression for the surface gradient we can estimate the effect of the interferometer's spatial frequency response on the overall magnitude of the reconstructed wavefront.   The fractional reduction in the magnitude of the thermal deformation due to spatial filtering from the interferometer was estimated as 
\begin{equation} \label{eq:spatialfilteringeffect}\frac{ h_{model}(0)-h_{measured}(0) }{ h_{model}(0) }\approx 1-\frac{1}{\phi(0)} \int_{-w}^{0}\frac{(\phi(r+s/2) - \phi(r-s/2))}{s} dr\end{equation} where the surface deformation is evaluated over one Gaussian radius of the pump beam - an approximation that is necessary because the analytical model being used assumes an infinite mirror radius with an infinite surface displacement.  The finite difference of points on the calculated surface profile for our shearing distance of $s=860\;\mathrm{\mu m}$ is taken at discrete points separated by $dx=9.8\;\mathrm{\mu m}$ representing the resolution of our image sensor.  We then subtract the mean value and calculate the cumulative sum, scaling the results by a factor of $dx/s$, which reproduces the original surface profile filtered by the spatial frequency response of our shearing interferometer.  This processing effectively smoothes the reconstructed surface resulting in a reduction in the measured peak deformation which is plotted in figure~\ref{fig:spatial_filtering} as a function of  $s/w$, the ratio of the shear to the Gaussian beam width of the heating beam.  For this experiment $s/w=0.91$ resulting in the measured peak deformation being reduced to 90\% of the original peak height.  Thus we can multiply our measured values by a factor of 1.11 to account for the spatial filtering. 

For the functional form for the thermal deformation given in \cite{Lu:07}, $l_c=0.94 w$, thus we have $s/l_c$= 0.96 so the shear is close to an optimum value.

\section{Measurement of a thermally induced mirror distortion}
\begin{figure}[h] `
\center
\includegraphics[width=.7\textwidth]{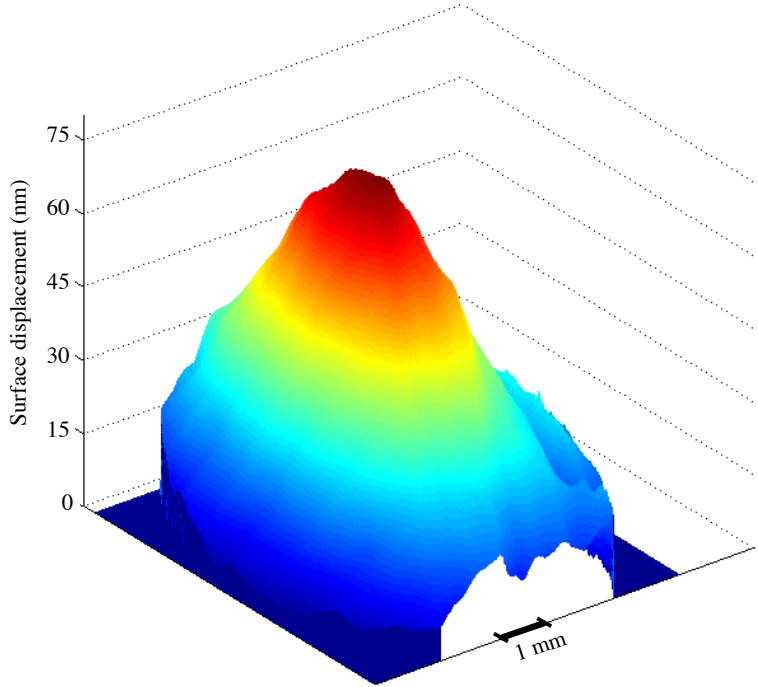}
\caption{\label{fig:heightmap}Thermal deformation measurement (in nm) from the shearing interferometer.  A peak height of 72 nm was observed prior to accounting for the spatial filtering of the shearing interferometer, corresponding to a height of 80nm when the filtering is taken into account. }
\end{figure}

The shearing interferometer was used to measure the thermal deformation of a gold mirror on a BK7 substrate.  A pump beam at 405 nm with a measured power of 78.9 mW illuminated the gold mirror at near normal incidence, with 19.7 mW reflected. 
With the transmission negligible this gives 59.2 mW of absorbed power.  The pump laser has a Gaussian beam profile with a Gaussian radius of 0.95 mm.  The thermal deformation corresponding to the reconstructed wavefront before accounting for the spatial filtering inherent in the shearing interferometer has a peak of 72 nm and is shown in figure ~\ref{fig:heightmap}, when scaled by a factor of 1.11 to account for the limited spatial frequency response at s/w=0.91 shown in Figure~\ref{fig:spatial_filtering} the corrected peak value is 80 nm.

 The magnitude of the deformation was  calculated to be 81 nm which is within 2\% of our measured value.  To further validate the results of the shearing interferometer, the thermal deformation of the test mirror was independently measured using a white-light Michelson interferometer, illuminated by a red LED with a center wavelength of $623\pm3$ nm.  The reference mirror of the interferometer was tilted to produce fringes on the interference pattern, which was recorded by the camera both with and without the test mirror being subjected to the pump beam that produced the thermal deformation.  The largest non DC peak of the 2-dimensional spatial Fourier transform was found, corresponding to the dominant spatial frequency $f_m$ of the interference fringes.  A superGaussian bandpass filter (n=8) centered at $f_m$ with a bandwidth of 25 spatial frequency units (one spatial frequency unit was $2\pi/640 \;\mathrm{rad/pixel}$) 
filters the data.  The Fourier transform was then down shifted by $f_m$, a process analogous to demodulation of a modulated time varying signal. The inverse Fourier transform of this filtered and shifted data was  computed.  This process, described in \cite{brooks2007} is analogous to the the transmission of a time varying signal by the process of modulation and demodulation.

This procedure was done for 2 different amounts of mirror tilt, corresponding to two different modulation frequencies (of 5 and 7 fringes over the field of view), to produce equivalent deformation maps.  Additionally the deformation of the mirror was measured by aligning the reference mirror to eliminate the tilt between the interfering wavefronts giving a uniform spatial profile of the interference pattern.  The interference pattern was recorded with and without the thermal deformation and the intensity of each pixel was compared to the maximum and minimum intensities seen as the reference mirror scanned through a full fringe so that the phase of the interference term could be deduced.  The wavefronts were reconstructed and the difference was taken to produce a deformation map of the mirror surface.  The 3 measurements of the mirror deformation with the white light interferometer yield a magnitude for the surface deformation of $75\pm13$ nm consistent with the values found by calculation and from measurements with the shearing interferometer.

\section{Applications}
While relative measurements of the magnitude of an arbitrary deformation are possible, accurately measuring the {\em absolute} magnitude of a deformation requires using a shear $s$ that is significantly smaller than the minimum length scale of interest, or requires prior knowledge of the shape of the deformation so that the effect of spatial filtering can be computed independently of the measurement.  Thermo-elastic surface deformations of an optic illuminated by a laser beam, such as that described in this paper, meet the latter criteria.  Laser interferometer based gravitational wave detectors have optics that are exposed to high power laser radiation producing thermal deformations that must be sensed and controlled for operation of the instrument at the design sensitivity \cite{LIGO:TCS}.  
An advanced LIGO intermediate test mass with radius of curvature of 1934m and a beam spot radius of 5.5cm has a sagitta measured over the beam spot of 782 nm \cite{Harry2010}. The theoretical sensitivity of this interferometer given in  equation~\ref{eq:hmin} with a Helium-Neon laser and an 8-bit imaging sensor ($\lambda=632.8$ nm, and $n=8$) allow this to be measured to 0.2 nm, equivalent to a change in the radius of curvature of the mirrors by 0.5 m, well within the required 20m accuracy \cite{Flanigan2008}.

\section{Conclusion}
We have presented a shearing interferometer using polarization control and spatial walk-off in a birefringent crystal to generate shear.  This configuration requires no critical alignment and  has excellent common-mode noise rejection.  We have demonstrated the ability to detect the wavefront distortion produced by a thermo-elasetic deformation in an optic heated by absorption of radiation from a laser beam.  Our instrument uses post-processed video of modulated interferograms to interpret the wavefront distortion.  We leave the development of software to process interferograms in real time as future work.  Our instrument has a sensitivity to surface gradients as small as $5.8\times 10^{-6}$, which allows a sensitivity of 3.8 nm of thermo-elastic deformation from heating by a Gaussian laser beam.  We have shown the sensitivity is optimized when the size of the shear is comparable to the characteristic length scale of the distortion being probed, and that this optimized displacement sensitivity is scale invariant.  We have shown how the effect of spatial filtering by the shearing interferometer can be calculated for a known distortion profile, and computed the magnitude of this effect for a thermo-elastic deformation of a mirror heated by a Gaussian beam as a function of the amount of shear.  Our measurement of the thermo-elastic deformation of a mirror is consistent with an analytical model and with an independent measurement using a white-light Michelson interferometer.

\end{document}